\documentclass[acus]{JAC2003}
\usepackage{graphicx}
\usepackage{multirow}
\usepackage{booktabs}
\usepackage{color} 
\usepackage{colortbl} 
\usepackage{subfigure} 
\usepackage{float}
\usepackage{placeins}

\begin{document}
\title{HIE-ISOLDE HIGH BETA CAVITY STUDY AND MEASUREMENTS}

\author{A.~D'Elia\thanks{Alessandro.Delia@cern.ch} and R. ~M. ~Jones, Cockroft Institute, UK, University of Manchester, Manchester, UK\\ M. ~Pasini, CERN, Geneva, Switzerland\\ and Instituut voor Kernen Stralingsfysica, K. U. Leuven, Leuven, Belgium}

\maketitle

\begin{abstract}
The upgrade of the ISOLDE machine at CERN foresees a superconducting linac based on two gap independently phased Nb sputtered Quarter Wave Resonators (QWRs) working at 101.28MHz and producing an accelerating field of 6MV/m on axis. A careful study of the fields in the cavity has been carried out in order to pin down the crucial e-m parameters of the structure such as peak fields, quality factor and e-m power dissipated on the cavity wall. A tuning system with $\approx$200kHz frequency range has been developed in order to cope with fabrication tolerances. In this paper we will report on the cavity simulations. The tuning plate design will be described. Finally the frequency measurements on a cavity prototype at room temperature will be presented.
\end{abstract}

\section{INTRODUCTION}

The upgrade of the ISOLDE machine will consist in boosting the energy of the machine from 3MeV/u up to 10 MeV/u with beams of mass-to-charge ratio of 2.5$\leq$A/q$\leq$4.5 and in replacing part of the existing normal conducting linac~\cite{matteo}. 

The new accelerator will employ superconducting QWR's based on Nb sputtered on Cu substrate technology~\cite{giulia} with a working frequency of 101.28MHz and with a nominal accelerating field of 6MV/m on beam axis. 

Two cavity geometries, {\em low} and {\em high} $\beta$, will be used to cover the whole energy range. As the first part of the upgrade will consist in the realization of the high energy section, the R\&D effort have been focused on the study and production of a prototype of the {\em high} $\beta$ cavity. 

The electromagnetic study and realization of the cavity prototype has been carried out in different steps. The definition of the cavity main parameters derives directly from beam dynamics studies (beam aperture, gap to gap distance, RF field asimmetry)~\cite{matthew}, from the upstream linac (RF frequency) and from manufacturing technique (Nb sputtering). Given the above constraints, the electromagnetic study has been performed in order to minimize the surface peak electric and magnetic field, maximizing the $R_{sh}/Q$ and the $g$ factor. In addition a particular attention has been payed to the study of the frequency sensitivity of the different geometric parameters, in order to evaluate a suitable tuning range. During the fabrication of the cavity prototype, after each significant step (machining, welding, deep pressing) an RF measurement has been performed and results are reported after in the paper. Given the limited frequency range of the tuner ($\approx$200kHz), the final dimension of the external conductor has been set only after the frequency measurement performed with the cavity still at the numerical control machine.       

\begin{figure}[htb]
\centering
\subfigure
[the cavity during the cleaning process] 
{\includegraphics*[width=40mm]{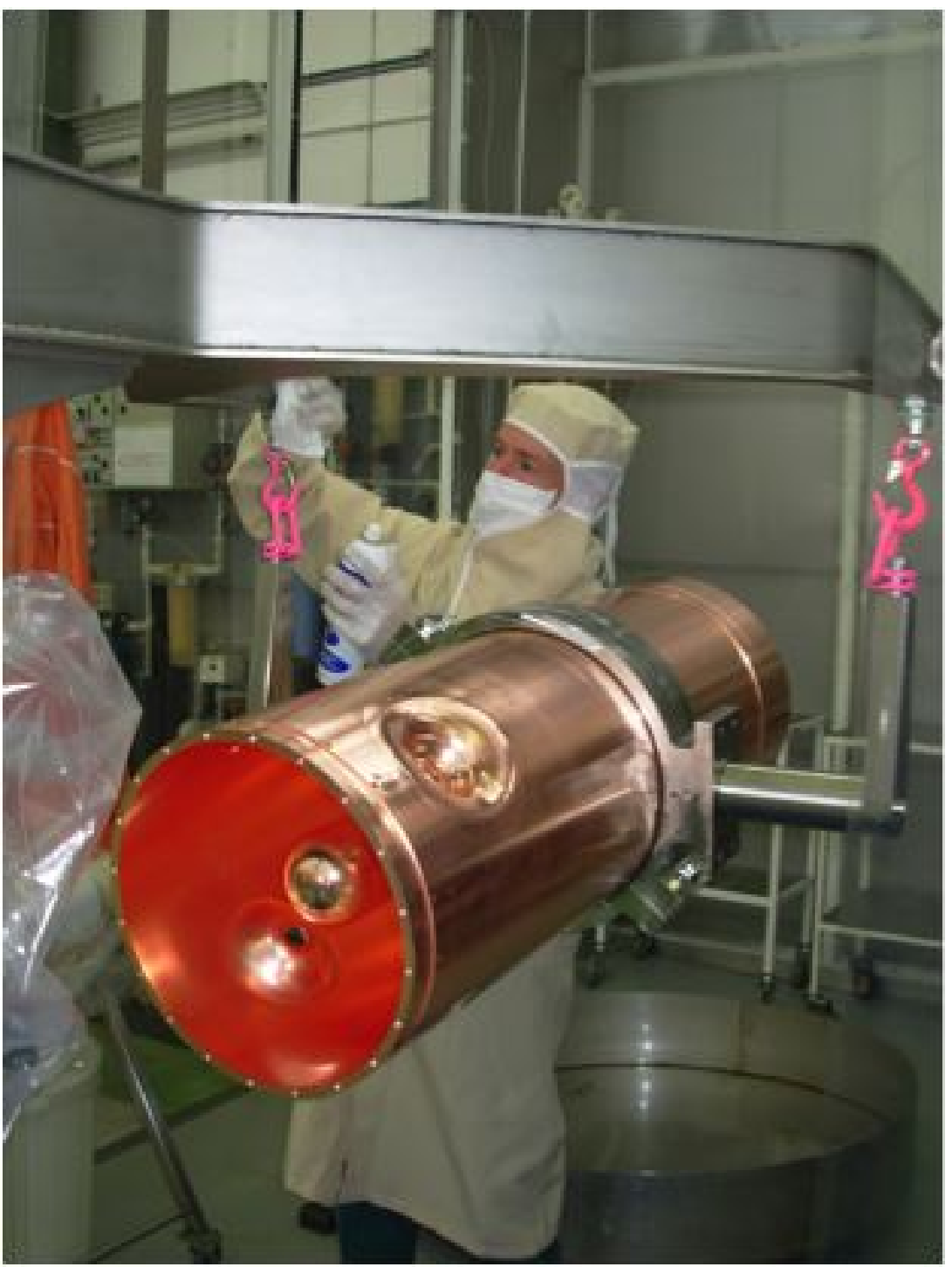}\label{a}}
\subfigure
[internal view of the cavity] 
{\includegraphics*[width=40mm]{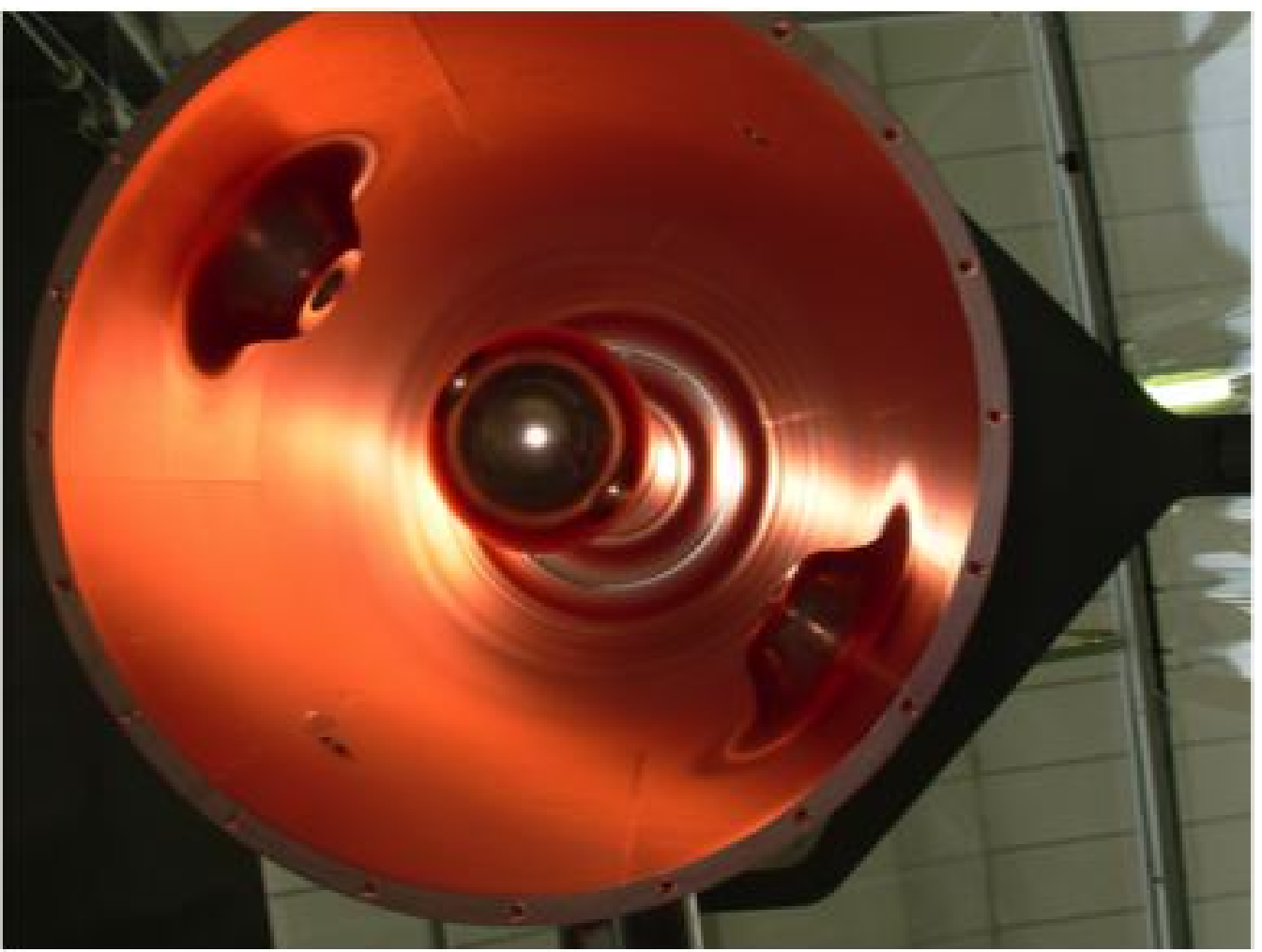}\label{b}}
\caption{The prototype of the high $\beta$ cavity produced at CERN.\label{cavity_cern}}%
\end{figure}

Figure~\ref{cavity_cern}  shows the prototype built at CERN and, presently, under sputtering tests. 

In the following we will show the main electromagnetic parameters for the optimized geometry of the cavity and we will compare them to the ones of similar structures as in TRIUMF and SPIRAL2. The tuner system will be presented. Finally the results of the frequency tests at room temperature at different steps of the machining will be reported. 

\section{ELECTROMAGNETIC SIMULATIONS}

The electromagnetic design of the cavity aims to minimize the surface peak fields, both electric and magnetic. Nb sputtering process will reduce the number of the useful possible shapes of the cavity: between the all possible geometries, the better suitable for Nb sputtering has to be chosen. Therefore the optimum is a compromise between RF performances and sputtering requirements~\cite{giulia}. 

In particular beam ports on the external conductor have been shaped in order to avoid any hidden edges on surface to the Nb cathode. Similarly, the region of the maximum magnetic field has been rounded in order to have a better homogeneity of the Nb film and this shape gives also a regular surface current paths minimizing the magnetic fields on these positions. 

\begin{figure}[htb]
\centering
\includegraphics*[width=80mm]{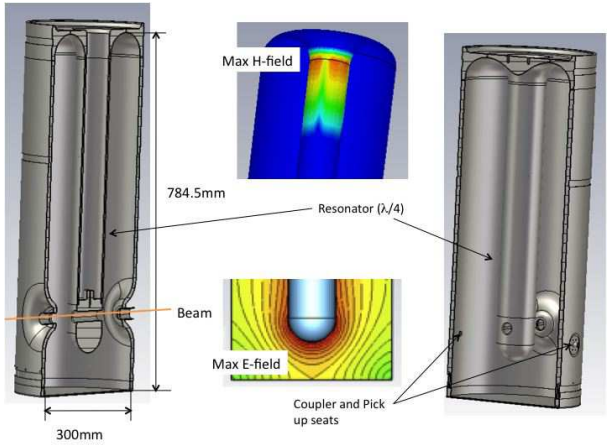}
\caption{Cut-view of the cavity showing the internal geometry and also coupler and pick-up seats; in the middle, the regions of the maximum electric and magnetic fields are shown.}
\label{cavity}
\end{figure}

The electric peak field, located at the bottom part of the resonator antenna, can be varied by changing the distance of the bottom part of the antenna from the bottom plate of the cavity (tipgap). The tipgap has been chosen of 70mm but the cavity, at the beginning, has been manufactured with a tipgap of 90mm. This gives the opportunity of tune up the frequency at the end of the machining procedures if needed. The simulations presented in the following have been carried out with the nominal value of 70mm. Figure~\ref{cavity} shows a cut-view of the CAD model and the positions of the maximum electric and magnetic field. 

\begin{figure}[htb]
\centering
\includegraphics*[width=75mm]{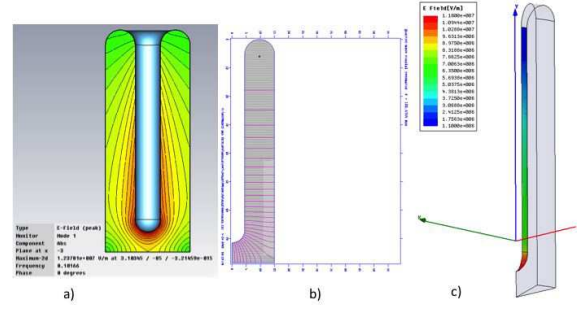}
\caption{Models used for simulations: a) MWS, b) SUPERFISH, c) HFSS.}
\label{comparison}
\end{figure}

Because of the cavity geometry is not azimuthal symmetric, 3D electromagnetic codes need to be used to evaluate its RF properties. In order to pin down reliable results, a {\em calibration} of the simulation tools have been performed. A simpler geometry without beam ports has been designed in order to get a comparison between Ansoft HFSS\copyright~\cite{hfss}, CST Microwave\copyright~\cite{mws} and POISSON SUPERFISH (Fig.~\ref{comparison}). For this kind of structures, we consider the data from SUPERFISH as the reference one.

\begin{table*}[!htbp]
\caption{Comparison table.}\label{comp_table}

\centering

\renewcommand\arraystretch{1}% (MyValue=1.0 is for standard spacing)

\small
\begin{tabular}{cccccc}
\hline
& \bf Superfish & \bf CST & \bf HFSS $\begin{array}{c} \bf CST-SF\\\%\end{array}$ &$\begin{array}{c} \bf HFSS-SF\\\%\end{array}$ \\ 
\hline
$\begin{array}{c} \bf Frequency\\\mbox{\bf (MHz)}\end{array}$ & 101.674 & 101.666 & 101.674 & - & -\\ 

$\begin{array}{c} \bf H_{peak}\\\mbox{\bf (kA/m)}\end{array}$ & 16.711 & 16.733 & 16.763 & 0.1 & 0.3\\ 
 
$\begin{array}{c} \bf E_{peak}\\\mbox{\bf (MV/m)}\end{array}$ & 11.38 & 11.57 & 11.6 & 1.7 & 1.9\\

$\bf Q_{0}$ & 11795 & 11844 & 11746 & 0.4 & -0.4\\
\hline 
\end{tabular}

\end{table*}

Considering the azimuthal symmetry of this simpler structure, only one quarter has been simulated in CST and the mesh refinement has been used. The higher step in the refinement corresponds to about 1.000.000 of meshcells. In HFSS, only a slice of one sixteenth has been simulated. The results have been obtained by considering a surface refinement of the meshes of 2$\mu$m, the total number of tetrahedra is about 110.000. The mesh stepsize for SUPERFISH is of 2.5mm (finer meshing does not change the results). 

\begin{figure}[htb]
\centering

\subfigure
[Max E-field] 
{\includegraphics*[width=90mm]{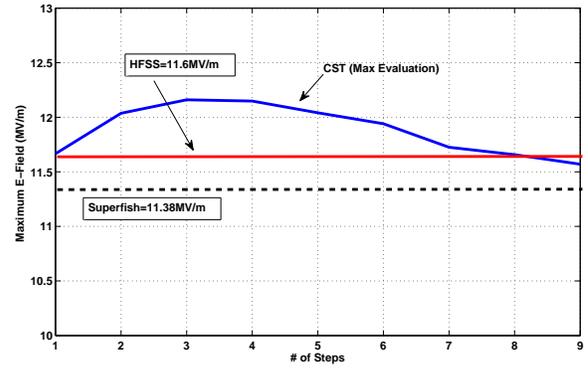}\label{a}}
\subfigure
[Max H-field] 
{\includegraphics*[width=90mm]{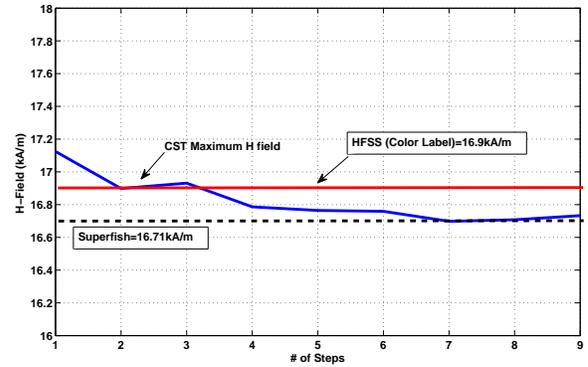}\label{b}}
\caption{Peak surface fields.\label{field}}%

\end{figure}

The comparison terms have been the maximum electric and magnetic field and the $Q_0$. The results for peak fields are shown in Fig.~\ref{field}. 

The fields are normalized to give 1J stored energy in the cavity (CST normalization). The plots show consistent results between the three codes. A summary of the results, included $Q_{0}$ values are listed in Table~\ref{comp_table}. 

When the beam ports are added to the structure, the results are expected to stay consistent with the ones found previously. For this structure, Superfish is no longer possible to use and then the results are extracted only from HFSS and CST. Actually $E_{peak}$ and $H_{peak}$ are consistent between HFSS and CST and also with the previous plots, but the value of $Q_{0}$ found by CST is overestimated, about 14000, and not consistent with the HFSS simulations, about 11700. Probably this is due to some geometry problem in the CST model around the noses. In fact, when the racetrack shape of the noses is changed to a simpler circular geometry, the value becomes compatible with other simulations.  
 
Finally, the electromagnetic cavity parameters are shown in Table~\ref{par} in comparison with TRIUMF and SPIRAL2.

\renewcommand{\thefootnote}{\alph{footnote}}

\begin{table}[!h]
\caption{Cavity parameters in comparison with TRIUMF and SPIRAL2.}\label{par}
\centering
\begin{minipage}{0.47\textwidth}
\renewcommand\arraystretch{1}% (MyValue=1.0 is for standard spacing)
\small

\begin{tabular}{c% 
>{\columncolor{yellow}}c% 
c% 
c}

\hline
 & \bf ISOLDE & \bf TRIUMF & \bf SPIRAL2\\
\hline
Frequency (MHz) & 101.28 & 141.4 & 88\\ 
$\beta$ (\%)& 11.4 & 11.2 & 12\\ 
$E_{acc}$ (MV/m) & 6 & 6 & 6.5\\
$L_{norm}$ (mm) & 300 & 180 & 410\\ 
$E_{peak}/E_{acc}$ & 5.4 & 4.9 & 4.9\\

$\begin{array}{c} B_{peak}/E_{acc}\\$[G/(MV/m)]$\end{array}$ & 96 & 99 & 90\\ 
   
$R_{sh}/Q_{0}$ ($\Omega$) & 554 & 545 & 518\\
g=$R_{s}\cdot Q_{0}$ ($\Omega$) & 30.34 & 25.6 & 37.5\\
$P_{cav}$ (W) & 7 & 7\footnote{Measurements on a cavity prototype showed results exceeding the design parameters: $Q_{0}$=7$\cdot 10^{8}$ for $E_{acc}$=8.5MV/m with $P_{cav}$=7W~\cite{Qtriumf}.} & 10\footnote{Measurements on a cavity prototype showed: $Q_{0}$=$10^{9}$ for $E_{acc}$=6.5MV/m with $P_{cav}$=10W~\cite{Qspiral2}.}\\
\hline 
\end{tabular}
\end{minipage}
\end{table}

\section{THE TUNER PLATE}

\begin{figure}[htb]
\centering
\includegraphics*[width=80mm]{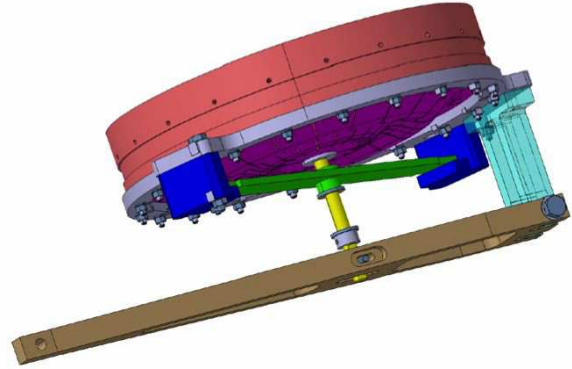}
\caption{Tuner plate with its actuator.}
\label{tuner}
\end{figure}

For the tuning system it has been decided to follow the concept that has been developed at TRIUMF~\cite{tuner1}. An oil-can shaped diaphragm of CuBe has been hydroformed with a pressure up to 120 bar. All radial slots necessary for the elongation and contraction of the diaphragm are performed with a laser beam. The same plate can be mounted directly on the low $\beta$ cavity or welded to a flange in the case of the high $\beta$ cavity. The actuator is designed to have no backlash. A pictorial view of the tuner is presented in Fig.~\ref{tuner}.

The useful stroke of the tuner plate is of 20mm. From the {\em manufacturing} position the plate can be pushed up towards the central resonator of 5mm (position +5) and down, in the other direction, of 15mm (position -15).   In Table~\ref{tuner_res} the results of the simulations are listed for the nominal value of the tipgap of 70mm and for a tipgap of 90mm. 

\begin{table}[htbp!]
\caption{Simulated frequency values of the tuning plate when all up (Position +5) or all down (Position -15); in yellow the values for the nominal tipgap value.}\label{tuner_res}
\centering
\renewcommand\arraystretch{1.5}% (MyValue=1.0 is for standard spacing)

\begin{tabular}{|c|%
>{\columncolor{yellow}}c|c|}
\hline
 & \bf Tipgap 70mm & \bf Tipgap 90mm\\ 
\hline
\bf Position +5 & 100.684 MHz & 101.235 MHz\\ 
%\hline
%$\bf\Delta_{tipgap}$ & \multicolumn{2}{c|} {27.55 kHz/mm}\\ 
 \hline
\bf Position -15 & 100.929 MHz & 101.339 MHz\\ 
%\hline
%$\bf\Delta_{tipgap}$ &  \multicolumn{2}{c|} {20.5 kHz/mm} \\ 
\hline
$\bf\Delta_{tuner\, plate}$ & 12.25 kHz/mm & 5.2 kHz/mm\\ 
\hline 

\end{tabular}
\end{table} 

The coarse range of the tuner plate for tipgap=70mm is foreseen to be of 245kHz for a moveable range of 20mm giving an average $\Delta_{tuner plate}\approx$12.25kHz/mm. The value for tipgap=90mm is of 5.2kHz/mm and the correction is less efficient as expected (coarse range of only 140kHz). However, both coarse ranges largely cope with frequency detuning coming from machining errors, listed in Table~\ref{err}, considering that mechanical tolerances are in the order of the tenth of mm. 

\begin{table}[h]
\caption{Simulations results of the frequency detuning due to the main possible machining errors.}\label{err}

\centering

\renewcommand\arraystretch{1}% (MyValue=1.0 is for standard spacing)
%\small
\begin{tabular}{cc}
\hline
\bf Type of error & $\bf\Delta_{freq}$ \bf (kHz/mm)\\

\hline
Cavity diameter & $\approx$65\\ 

Resonator length & $\approx$160\\ 
 
Nose length & $\approx$50\\

\hline
\end{tabular}

\end{table} 

Another important aspect to take into account is the radiation pressure acting on the tuner plate. Because of the high number of cuts and the reduced thickness of the diaphragm, a strong force could irreversibly modify the tuner shape leading to an unrecoverable frequency shift. The radiation pressure can be calculated by evaluating both electric and magnetic field on the plate surface by the following equation:

\[
 P=\frac{1}{4}\left(\mu_{0}|\bf{H}|^{2}-\epsilon_{0}|\bf{E}|^{2}\right)
\]

\begin{figure}[htb]
\centering
\includegraphics*[width=90mm]{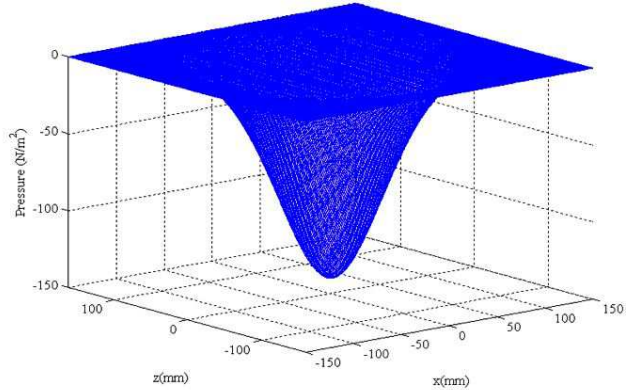}
\caption{Radiation pressure on a flat plate.}
\label{press_plot}
\end{figure}

The result is shown in Fig.~\ref{press_plot} in the case of a flat plate: the total force is quite low, equal to -1.77N.  

\section{FREQUENCY MEASUREMENTS}

The working frequency of the cold cavity has to be 101.28MHz. Taking into account the scaling factors due to the superconducting mode of operation (shortening of the length of the central resonator, skin depth variation, etc...) and due to the vacuum-air environment change, the {\em hot} frequency at room temperature in air should be $\approx$100.900MHz. 

As described before we have left the value of the tipgap as a {\em free} parameters in order to compensate possible errors coming from the machining process. During the welding of the top part of the cavity, a problem occurred and the result has been a change of the length of the internal resonator, shortened of about 0.4mm, with a foreseen increasing of the frequency of about 65kHz. The measured variation is of 77kHz. In Table~\ref{Q_freq} the measured values before and after welding the inner conductor to the outer are shown. It is also shown the value of Cu $Q_{0}$ which is significantly improved after welding as expected. Furthermore the measured value is consistent with simulations giving a value of about 11700.

\begin{table}[!h]
\caption{Measurements before and after welding the inner conductor to the outer with tipgap=90mm.}\label{Q_freq}

\centering
\renewcommand\arraystretch{1}% (MyValue=1.0 is for standard spacing)
\small
\begin{tabular}{cccccc}
\hline
 &\bf Before welding  & \bf After welding\\ 
\hline
 Frequency & 101.147 & 101.224\\ 

$Q_{0}$ & 5908 & 11380\\ 
\hline 
\end{tabular}

\end{table} 

The last step is to finally define the tipgap size. An intermediate cut of the bottom of the cavity has been defined of 75mm and directly in the mechanical atelier a set of measurements with two different coupler insertion lengths have been done to cross-check the simulated values. The results are shown in Table~\ref{measurements}: the values both of the simulations and of the measurements have been properly scaled (values in bold) to get a comparison. The effect of the lengthening of the central resonator can be accepted and the cut has been done at the nominal tipgap length of 70mm.  

\begin{table}[h]
\caption{Set of measurements for different cut of the tipgap and different coupler insertion; bold values are scaled to get a comparison.}\label{measurements}
\centering
\footnotesize
\renewcommand\arraystretch{1.1}% (MyValue=1.0 is for standard spacing)

\begin{tabular}{|c|c|c|}
\hline
 
  \multicolumn{3}{|c|} {\bf Coupler\, in=64mm\,\,and \bf Pickup\, in=22mm} \\
 
 \hline

% &  & Measurement \\ 

Tipgap 90 & $\begin{array}{c}\mbox{Simulation} \\\mbox{101.233MHz}\\(\mbox{-32kHz air})\\ \mbox{\bf 101.201MHz}\end{array}$ & $\begin{array}{c}\mbox{Measurement} \\ \mbox{101.246MHz}\\(\mbox{-77kHz res})\\ \mbox{\bf 101.169MHz}\end{array}$\\

\hline 
 
% Tipgap 75 & Simulation & Measurement\\ 
 
Tipgap 75 &  $\begin{array}{c} \mbox{Simulation} \\\mbox{101.013MHz}\\(\mbox{-32kHz air})\\ \mbox{\bf 100.981MHz}\end{array}$ & $\begin{array}{c} \mbox{Measurement} \\\mbox{101.000MHz}\\(\mbox{-77kHz res})\\ \mbox{\bf 100.923MHz}\end{array}$ \\
 
 \hline

%Tipgap 70 & Simulation & Measurement \\ 

Tipgap 70 & $\begin{array}{c}\mbox{Simulation} \\ \mbox{100.899MHz}\\(\mbox{-32kHz air})\\ \mbox{\bf 100.867MHz}\end{array}$ & $\begin{array}{c}\mbox{Measurement} \\ \mbox{100.916MHz}\\(\mbox{-77kHz res})\\ \mbox{\bf 100.839MHz}\end{array}$ \\ 
 
 \hline 
 
  \multicolumn{3}{|c|} {\bf Coupler\, in=22mm\,\,and \bf Pickup\, in=22mm} \\

\hline

%Tipgap 90 & Simulation & Measurement \\ 

Tipgap 90 & $\begin{array}{c} \mbox{Simulation} \\\mbox{101.410MHz}\\(\mbox{-32kHz air})\\ \mbox{\bf 101.378MHz}\end{array}$ &$\begin{array}{c}\mbox{Measurement} \\ \mbox{101.483MHz}\\(\mbox{-77kHz res})\\ \mbox{\bf 101.406MHz}\end{array}$\\

\hline 
 
 %Tipgap 75 & Simulation & Measurement \\ 
 
 Tipgap 75 & $\begin{array}{c} \mbox{Simulation} \\\mbox{101.191MHz}\\(\mbox{-32kHz air})\\ \mbox{\bf 101.159MHz}\end{array}$ & $\begin{array}{c} \mbox{Measurement} \\\mbox{101.240MHz}\\(\mbox{-77kHz res})\\ \mbox{\bf 101.163MHz}\end{array}$\\
 
 \hline

%Tipgap 70 & Simulation & Measurement \\ 

Tipgap 70 & $\begin{array}{c} \mbox{Simulation} \\\mbox{101.083MHz}\\(\mbox{-32kHz air})\\ \mbox{\bf 101.051MHz}\end{array}$ & $\begin{array}{c}\mbox{Measurement} \\ \mbox{101.150MHz}\\(\mbox{-77kHz res})\\ \mbox{\bf 101.073MHz}\end{array}$ \\ 

\hline
\end{tabular}
\end{table} 

Finally the measurement with the tuning plate has been done. The value is shown in Table~ \ref{plate_meas} in comparison with simulation results opportunely scaled and goal frequency and all the values are fully compatible.

\begin{table}[!htbp]
\caption{Frequency measurement at room temperature with tuner plate in rest position.}\label{plate_meas}

\begin{center}

\renewcommand\arraystretch{1}% (MyValue=1.0 is for standard spacing)
\small
\begin{tabular}{|c|c|}
\hline
 \rowcolor{yellow} Measured frequency &\bf 100.885 MHz \\ 
\hline
 Simulation & $\begin{array}{c} \mbox{100.816 MHz}\\(\mbox{-32kHz air})\\ (\mbox{+77kHz res})\\\mbox{100.861MHz}\end{array}$\\ 
\hline
\rowcolor{yellow} Goal frequency & $\bf\approx$\bf 100.900 MHz\\ 
\hline 
\end{tabular}

\end{center}

\end{table} 

\section{CONCLUSION}

The high $\beta$ cavity for ISOLDE upgrade has been fully designed and built. The cavity parameters have been derived showing values comparable to other similar structures (TRIUMF and SPIRAL2). The foreseen $Q_{0}$ should be $6.6\cdot 10^{8}$ with a surface resistance $R_{s}$=46n$\Omega$ giving a power dissipation on the cavity wall of 7W.

A prototype tuner plate has been built. The total coarse range in simulation is of 245kHz for a total course of 20mm giving 12.25kHz/mm.

The frequency measurement at room temperature shows a perfect agreement with the designed frequency: the measured frequency is 100.885MHz, the design frequency at room temperature should be $\approx$100.900MHz (the simulations give 100.861MHz). 

Presently the cavity and the tuner plate are under sputtering tests.

\section{ACKNOWLEDGEMENTS}
Thanks to Cockroft Institute for the financial support.
Thanks to Philippe Perret who took charge of the mechanical design of the tuning plate.

\end{document}